\documentclass[review]{mandm}
\usepackage{array}
\usepackage[utf8]{inputenc}
\usepackage{mathabx}
\usepackage{graphicx}
\usepackage{hyperref}
\usepackage{adjustbox}
\usepackage{subcaption}
\usepackage{float}
\usepackage{placeins}
\usepackage{stfloats}
\usepackage{lineno}
\usepackage{amsmath}
\usepackage{xcolor}

\jourvolume{XX}
\jourissue{Y}
\jourpubyear{20ZZ}

\begin{document}

\sloppy

\title[Accepted for Publication, Microscopy \& Microanalysis]{Towards Augmented Microscopy with Reinforcement Learning-Enhanced Workflows}

\author[M. Xu et al]{Michael Xu$^1$,
  Abinash Kumar$^1$,
  and James M. LeBeau$^1$}

\affiliation{%
$^1$Department of Materials Science and Engineering, Massachusetts Institute of Technology, Cambridge, MA 02139, USA\\
  Corresponding Author: James M. LeBeau \email{lebeau@mit.edu}}

\begin{frontmatter}

\maketitle

\begin{abstract}
Here, we report a case study implementation of reinforcement learning (RL) to automate operations in the scanning transmission electron microscopy (STEM) workflow. To do so, we design a virtual, prototypical RL environment to test and develop a network to autonomously align the electron beam without prior knowledge. Using this simulator, we evaluate the impact of environment design and algorithm hyperparameters on alignment accuracy and learning convergence, showing robust convergence across a wide hyperparameter space. Additionally, we deploy a successful model on the microscope to validate the approach and  demonstrate the value of designing appropriate virtual environments. Consistent with simulated results, the on-microscope RL model achieves convergence to the goal alignment after minimal training. Overall, the results highlight that by taking advantage of RL, microscope operations can be automated without the need for extensive algorithm design, taking another step towards augmenting electron microscopy with machine learning methods.

\end{abstract}

\end{frontmatter}

\section{Introduction}

Continued advancement of (scanning) transmission electron microscope technology has produced an ever expanding repertoire of imaging and spectroscopy techniques with which to probe materials and their defects \citep{Ede_2021deeplearning}. Through this growing complexity, challenges are presented in the form of time and experience needed to become proficient at using the microscope. Even for experienced users, the data collection process often begins with potentially time-consuming alignments, finding a region of interest, setting up the sample alignment (tilting), and determining optimum experimental conditions. Consequently, relatively small volumes are typically studied within a reasonable time frame. Combined with a limited field of view, this can lead to poor representation of the statistical average of the bulk material. Improving the user experience and acquisition efficiency of the electron microscope therefore has significant implications for materials research and data significance.

Augmentation and automation of electron microscopy have the potential to address instrument complexity while improving reproducibility and precision of microscope control. For example, automated imaging, data processing, and analysis have already proven essential as part of the workflow for biological studies with cryo-electron microscopy (CryoEM) \citep{callawayCRYO}. In this rich ecosystem, many open source and commercial solutions have been developed, offering automated control, real-time sample and microscope feedback, and even live data streams for processing and analysis{\color{red} \citep{tegunov_warp, epu2, petascale}}. 

In contrast to biological microscopy, automation for materials science studies are complicated by far greater sample heterogeneity and a broader range of operating conditions. Structural and chemical analyses of catalysic nanoparticles, for example, are critical to understand  functionality. Acquiring statistically significant atomic resolution measurements of such particles is, however, challenging \citep{np_catalysis, surface_strain_NP}. Likewise, the investigation of short-range order or clustering, where local chemistry varies and has implications for electrical or mechanical properties, atomic resolution imaging and large area mapping are both necessary and time consuming for manual analysis \citep{kumar2021pmn, Li_Sheng_Ma_2019}. For doped materials such as semiconductors or solid-state systems, low concentrations of dopants often present a challenge for locating, imaging, and analysis \citep{gunawan2011dopant, yang_DL_dopant2021}. These issues are magnified with a beam-sensitive material, where juggling alignment of the sample and imaging/spectroscopy parameters can limit reproducibility. 

{\color{red} To increase statistical sampling with electron microscopy in materials science, incorporating flexibility and ``intelligence'' within control algorithms is critical \cite{Spurgeon_Ophus_datadriven}. Significant progress has been made in open-source hardware control \citep{serialem, pyjem, nionswift} and data acquisition and analysis \citep{atomai, stempy} that set the stage for these automated experiments. Further, the adoption of machine learning (ML) and general artificial intelligence (AI) in these software packages shows the interest and promise of these tools as potential solutions for intelligent algorithms \citep{deeplearningmatscireview, Ede_2021deeplearning}. When combined with control systems and analysis tools, ML has been demonstrated to guide autonomous decision-making within experiments \citep{Olszta_sparseauto, autoDKL, akersnature}.} Additionally, with recent examples of automated nanoparticle imaging \citep{autoEM}, structure identification \citep{temimagenet, Ziatdinov_defects_transformsseg_Kalinin_2017, trackingsideep}, chemical analysis \citep{Ragone_DLHEAFCN_2022, deeptomoedx}, the future of supervised and unsupervised learning in microscopy is promising. 

Advancements in reinforcement learning (RL), another paradigm of ML, have the potential to enhance data collection and microscope workflows. These algorithms have been demonstrated to achieve beyond-human performance in increasingly complex systems, such as high-skill games \citep{silver_go, starcraft}. In the context of electron microscopy, this family of algorithms can provide a robust approach to augmentation and automation for materials characterization through instrument control. For example, RL has already been shown to tackle specific optimization challenges, such as compressed sensing through partial scanning \citep{partial_scanning}, electron beam-induced atomic assembly \citep{vasudevan_ebeam}, and autofocus \citep{SEM_RL_autofocus}. The ability of these algorithms to interact with complex environments is apparent, yet the general design approach for systematic automation of the microscope using RL can benefit from further evaluation \citep{Kober_Bagnell_Peters_2013}. Moreover, by designing simulated environments and leveraging off-the-shelf algorithms, RL can breathe new life into standard microscope interfaces and functionality through augmentation of the imaging process.

In this work, we report a case study implementation of reinforcement learning to augment the STEM workflow by autonomously aligning the bright-field disk. While a brute-force approach to this task may involve iterating over shift values using gradient descent or grid search methods with calibrated deflections, we demonstrate the potential of RL-based microscope control using a simple, proof-of-concept task: centering the STEM bright-field disk to a target location, e.g.~centered to a detector. With insights that can be extended or generalized given {\color{red} current microscope control frameworks and} cross-task similarity and modularity, details of environment design, algorithm implementation, and hyperparameter optimization are discussed. We show the feasibility of offline design and tuning applied to online training and validation, achieving successful centering without prior knowledge of the action space (shift axes and step calibration). {\color{red} These results provide initial guidelines for RL environment design, complementing existing efforts to automate operation experimentation and offering insight into important considerations for RL-based microscope control. }




\section{Reinforcement Learning}

\subsection{Overview}

Reinforcement learning describes a family of algorithms in which an agent interacts with an environment and receives feedback in the form of an observation. Using the observation, the agent then selects another action in the environment and observes the following response, with the ultimate objective of maximizing a cumulative reward function. This reward is usually maximized by attaining a goal state, upon which the interaction period, or episode, ends \citep{Sutton_Barto_2018}. 

An example of the RL approach is seen in $Q$-Learning, where an agent learns to take discrete actions across a set of states to maximize its expected episode reward. After repeatedly interacting with the environment over many timesteps, the $Q$-value function, used to estimate the future reward of present actions and states, is updated, and future actions are chosen in order to maximize the total reward. In essence, the agent learns an accurate value function to estimate the future reward and guide decision making, as well as a successful action strategy, or policy, using this value function. 

Moreover, in RL environments with continuous actions or observations, such as in the control systems of electron microscopes, calculating the value function across all actions and states suffers from the curse of dimensionality. As a result, different RL agents have been developed to address the challenges of navigating nearly infinite states or actions. For example, actor-critic frameworks aim to separately learn a value estimator and action estimator, allowing states and actions to be learned separately and decreasing the number of samples needed to form a successful policy \citep{Sutton_Barto_2018}. These types of algorithms have been demonstrated to thrive in complicated environments, including robotic control or advanced physics simulators \citep{lillicrap_continuousRL, openaigym}. In particular, Soft Actor-Critic (SAC) is a type of actor-critic method that establishes a trade-off between state-action exploration and maximizing future rewards. The long-term expected reward (exploitation) and entropy (exploration) are balanced using an entropy coefficient in the value estimator, which can be automatically optimized during learning. \cite{Haarnoja2018}, for example, have shown the stable learning and sample-efficient behavior of SAC in a variety of virtual and real-world robotic control environments.

The microscope operation is also characterized by a large state space, and  can be defined as optimization of specific imaging conditions over a complex combination of parameters. Collecting data often requires continuous adjustments of several parameters (e.g. beam deflection, stage movement, etc.) with sparse feedback in the form of an image or value (e.g. diffraction pattern, screen current, etc.). Reinforcement learning thus offers a well-suited method for optimization of these challenges \citep{lillicrap_continuousRL}. Using this model, the available actions for an RL algorithm encompass those also accessible to the standard user, such as lens adjustments, and the states or observations can be either directly read or, like a human user, inferred from visual feedback. In addition, typical electron microscope operation can be broken down into smaller sub-tasks. These sub-tasks share many commonalities, such as centering a feature (the beam or diffraction pattern), maximizing a value (focus/image contrast, beam current, etc.), or `reading' a diffraction pattern (symmetry, thickness estimation, etc.). Consequently, an RL implementation may also be partitioned, reducing the complexity of any individual challenge. For example, centering of the bright-field disk or convergent beam electron diffraction (CBED) pattern is a procedure that has significance on STEM image formation, since proper alignment with respect to individual STEM detectors can impact the image intensities and contrast \citep{quant_detectorADF,Findlay_LeBeau_2013_detector}. As a result, this challenge presents an exemplary model to guide the development of simulated and on-microscope environments for RL-based control. 

\subsection{Design and Implementation}


The OpenAI Gym \citep{openaigym} template for environment design was used to develop the simulated and on-microscope RL environments in order to take advantage of its standardized API and RL agent compatibility. For alignment of the bright-field disk, vector inputs and outputs were chosen in lieu of complete images. The observation state included: difference between the measured center of mass and the goal position (dx, dy), True/False if either one of the residual coordinates are zero (dx=0, dy=0), and an arbitrary ``step size'' multiplier, to account for magnitude differences between shift values and on screen movement. The disk center of mass was calculated on a sub-pixel basis using the raw camera pixels and sub-sampled with a pre-specified number of bins for the RL environment. {\color{red} We note that the center of mass provides a robust method for measuring the position of the bright-field disk for most imaging doses, except when the intensity of the pattern approaches the level of noise due to camera limitations or improper gain and exposure correction. In our case, a total probe current of 10 pA corresponding to 0.0035 pA/px of the bright field disk yielded stable center of mass measurements, though this may be improved with optimized camera settings (gain, gamma, etc.). }

The environment was automatically reset to a random position on the screen when 1) the goal position was attained, 2) the beam was outside the threshold, or 3) the step limit was reached. During development, the environment complexity was increased from a simple 2D observation space and discrete action space to a continuous action space scaled to more closely match the microscope shift range available. Random offsets to the x-y axes rotation and random noise of the action were implemented to account for any non-linearity or noise in the actual system. 

The final stepwise reward, given by Equation \ref{eq:rewards}, was carefully tuned during initial testing of the environment. If the bright-field disk center of mass moves closer to/further from the goal position than in the previous step, the reward is $\pm$1, respectively. If the position is within a given tolerance from the goal in pixels, the reward for that terminal step is +100, and if the disk center of mass moves a certain threshold distance near the edge of the screen or the maximum number of steps in the episode is reached, the reward for the terminal step is -100. Additionally, a distance-based discount scaled to the observation space size was applied in order to further penalize movements away from the goal and offer a more informative reward. For total episode reward, the stepwise reward is summed across all steps within the episode. 

\begin{equation}
\begin{adjustbox}{max width = 100 mm}
    $R(s_t, a_t, s_{t+1}) = \begin{cases} 
          -1-\dfrac{d_{s_{t+1}}}{\textrm{size}} & d_{s_{t+1}} > d_{s_t}\\[7pt]
          +1-\dfrac{d_{s_{t+1}}}{\textrm{size}} & d_{s_{t+1}} \leq d_{s_t}\\[7pt]
          -100 & \textrm{no beam}\\
          100 & d_{s_{t+1}}< \textrm{tol.}
       \end{cases}
    $
\end{adjustbox}
\label{eq:rewards}
\end{equation}

Due to the continuous control aspects of electron microscopy, an implementation of the Soft Actor-Critic (SAC) algorithm from the Stable-Baselines3 Python package was used \citep{Stable-Baselines3}. To mitigate the issue of lengthy and costly on-instrument time needed for development and to expedite debugging and hyperparameter tuning, a simulated bright-field disk alignment environment was created with which RL experiments were performed. The simulated environment generated a similar image and output to what is received by from acquisitions of the phosphor screen FluCam of the microscope.

On-microscope tests of RL were performed with a probe aberration-corrected Thermo Fisher Scientific Themis Z 60-300 kV S/TEM system. A custom Python interface, based on the Universal Scripting Engine for TEM (USETEM), was developed to continuously read and acquire the gain- and exposure-corrected image stream from the phosphor screen FluCam for training and testing of RL-enhanced functionality \citep{LeBeau_Kumar_Hauwiller_2020}. Accessing this data feed is beneficial, as it offers a wealth of information regarding the microscope and sample states without needing more sensitive CCDs or CMOS sensors.

\begin{figure}
\centering
\includegraphics[width=3.2in]{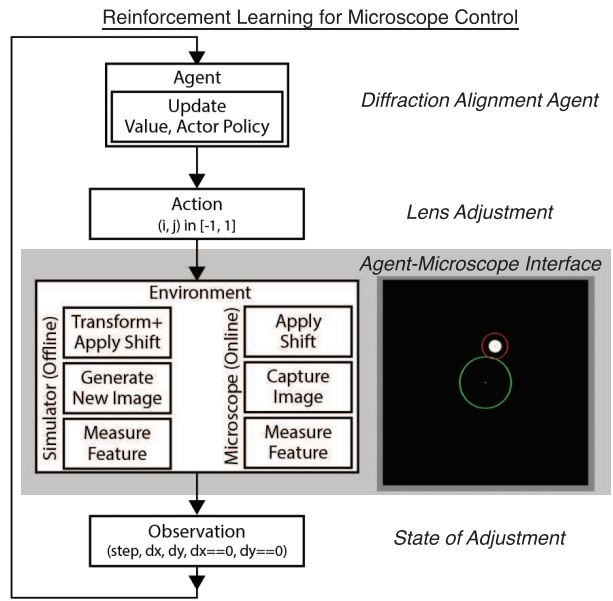}
\caption{Overview of the reinforcement learning process for diffraction alignment. The agent repeatedly samples and sends actions in the form of a diffraction shift, which in turn produces a simulated (offline) or measured (online) environment response that can be observed. The resulting observation is used to update the value and actor policies, which dictate future decision making. A custom Python program was developed to access the phosphor screen FluCam and send commands to the microscope. Image acquisitions from the camera are overlaid with the screen center (green) as well as the current image center of mass (red). }
\label{fig:flucamgui}
\end{figure}

The flow of the simulated and microscope environments is shown in Figure \ref{fig:flucamgui}. The RL agent interacted with these environments by selecting an action, receiving an observation and reward value, and updating its internal neural network weights to better estimate the rewards and best actions to take. In all, a total of 1,500 learning periods of 20,000 timesteps each were run in the simulated environment to measure the hyperparameter sensitivity of the RL algorithm. Reward design, environment sub-sampling, number of neurons, goal tolerance, and step size were evaluated. Total steps per episode and total episode reward describe the success and efficiency of algorithms and are used to evaluate this performance, with exponentially weighted averaging using a window of 25 points used to reduce noise for determining convergence. The resulting environment template and RL training script can be downloaded at https://github.com/LeBeauGroup/DiffAlignRL.

\section{Results \& Discussion}

\subsection{General Performance}

\begin{figure}
\centering
\includegraphics[width=3.2in]{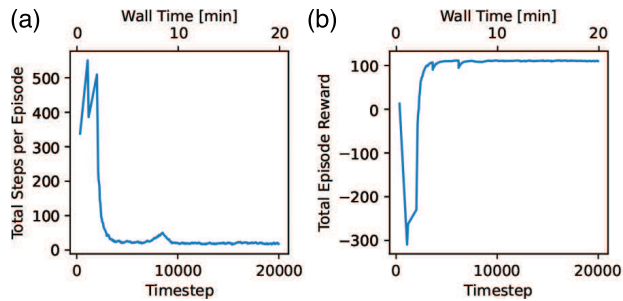}
\caption[name]{Convergence of an agent shown by the (a) total number of steps per episode and (b) cumulative episode reward as a function of elapsed timesteps (and corresponding wall time) in the simulated environment.}
\label{fig:genRLcurve}
\end{figure}

The definition of a successfully trained RL algorithm is when it reaches its goal state repeatedly, with the episode reward converging to a maximum. Here, the total steps per episode and total episode reward for a typical agent is shown in Figures \ref{fig:genRLcurve}a,b, respectively. With the definition of success, learning in the simulated bright-field disk alignment environment yielded successful agents well within the allotted 20,000 learning timesteps, achieving a maximum reward and minimum number of actions for correct alignment in many cases by 3,000-4,000 steps or a wall time of 3-4 minutes. 

Two distinct intervals comprising the learning curves are seen: an initial time span of low performance and a period of rapid improvement to final optimal performance. In the first 1,000-2,000 timesteps, the number of steps per episode increases to a maximum and the total episode reward reaches a minimum negative value. Very few episodes are seen in this interval, as the position of the bright-field disk meanders on the screen, rarely reaching the terminating conditions associated with going off screen or to the goal. However, this period is quickly followed by rapid improvement, in which episodes are completed in a fewer number of steps and the total episode reward increases. In this case, either more, correct alignments or faster movement off screen would minimize the number of steps and prevent accrual of highly negative rewards. 

\begin{figure}
\centering
\includegraphics[width=3.2in]{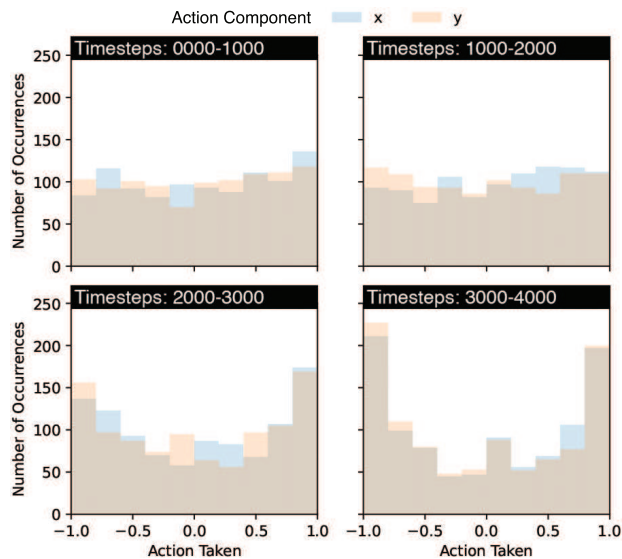}
\caption[name]{Evolution of actions taken with respect to learning progress in 1,000-timestep intervals for a successfully trained agent. }
\label{fig:genRLact}
\end{figure}

\begin{figure}
\centering
\includegraphics[width=3.2in]{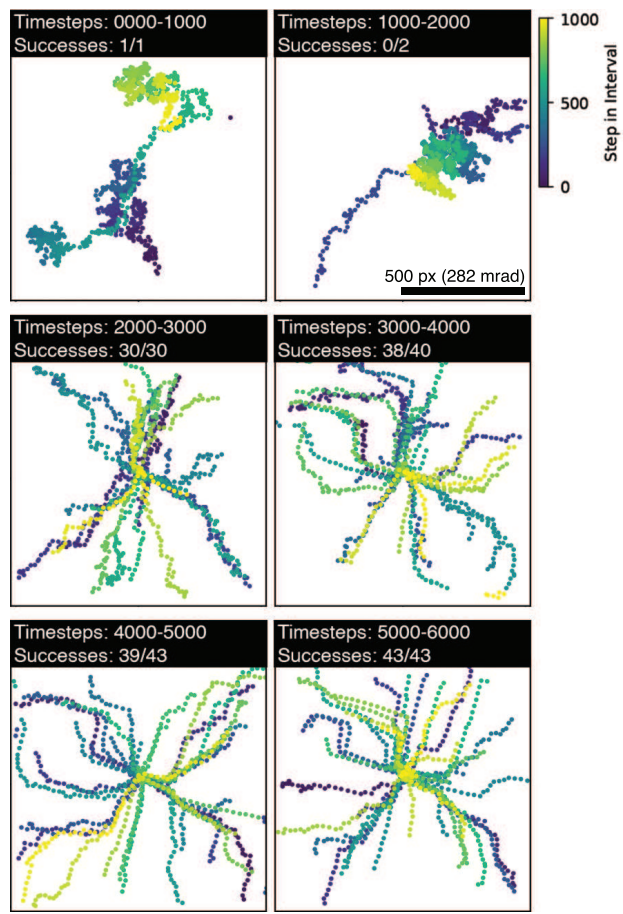}
\caption[name]{Beam positions separated into 1,000-timestep intervals for a typical successful agent in the simulated environment. The number of successes out of the total completed episodes in each interval is shown. }
\label{fig:genRLobs}
\end{figure}

To better understand this progress, action and observation states during specific intervals of this learning period are visualized in Figures \ref{fig:genRLact} and \ref{fig:genRLobs}. Initially, actions are sampled uniformly over the entire interval (-1, 1), with the bright-field disk frequently driven aimlessly or out of bounds, which is confirmed by looking the path of the disk within a given timestep interval. Over the course of learning, the distribution of actions shifts towards those of larger magnitudes as the agent becomes more efficient at moving towards the goal, with the number of successes increasing for a fixed timestep window (Figure \ref{fig:genRLact}). The corresponding paths of the beam show this as well, as both a higher number of total episodes and a greater number of successful alignments are achieved (Figure \ref{fig:genRLobs}). Eventually, a minimal number of timesteps (21.4 $\pm$ 10.4 for this experiment), equivalent to a wall time of under one second, is needed for the agent to successfully complete an alignment episode, or reach the goal condition from a typical initial starting point. These characteristics correspond to a transition from state-action sampling towards maximization of rewards. We also observe that a small increase in number of steps and two small drops in episode rewards are seen from Figure \ref{fig:genRLcurve}, which we attribute to continual adjustment of the action policy, or strategy, by exploration of states and maximization of rewards. The Supplemental Material Video shows the training progression for a successful agent. 

\subsection{Alignment Precision/Tolerance}

Accuracy and precision of bright-field disk centering using the RL algorithm is primarily affected by the pre-specified convergence conditions, or goal error tolerance, as well as the observation space sub-sampling, or binning, of the environment. These factors determine the threshold at which the disk is considered centered and the reduction in size of the two-dimensional ``grid'' in which the agent operates. Learning using several error thresholds and varying degrees of binning were performed to determine alignment accuracy using the distribution of final corrected positions in sub-pixel camera space, from which an expected error was calculated in terms of phosphor screen camera pixels and calibrated angular pixel size (mrad). 

\begin{figure}
\centering
\includegraphics[width=3.2in]{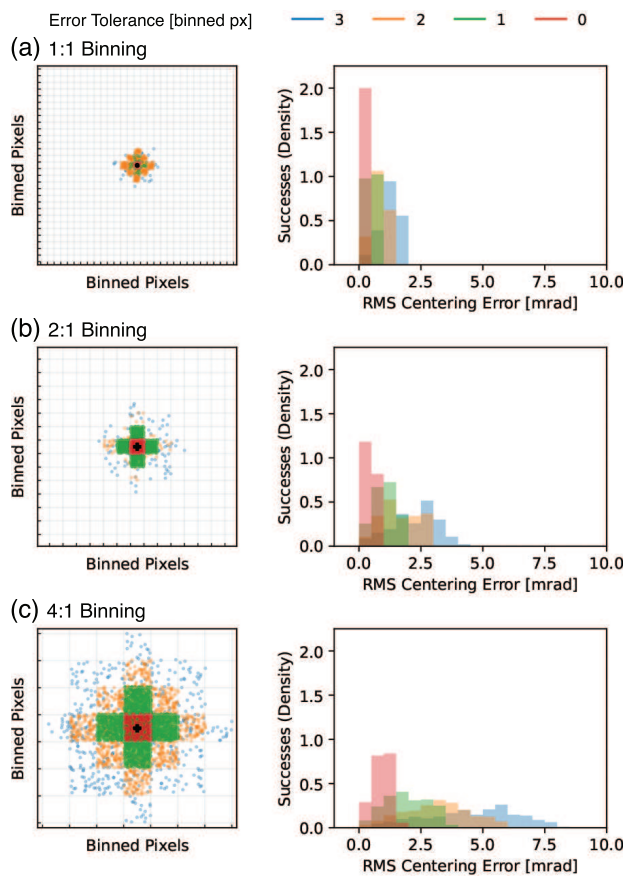}
\caption[name]{Final successful states from all simulated environment tests as a function of goal error tolerance and a) 1:1, b) 2:1, and c) 4:1 raw camera pixel binning, and corresponding distributions of root-mean-square (RMS) centering error in calibrated pixel size (mrad). The y-axis of RMS centering error plots is given as a density normalized to the total number of successes for each combination of error tolerance and binning degree.}
\label{fig:pixel_tol}
\end{figure}

With the calibrated pixel size of the phosphor screen camera of 0.565 mrad/px, the value of the goal state is given as a sub-pixel numerical value in camera space before environment binning. An optimally accurate algorithm is observed with no camera binning and 0 binned pixel tolerance, yielding an expected root-mean-square (RMS) centering error of less than half a camera pixel, corresponding to sub-0.5 mrad (Figure \ref{fig:pixel_tol}a). Additionally, the successful terminal states appear to be uniformly distributed over one pixel, with a mean square error (MSE) of approximately 0.137 mrad. With additional binning, the larger environment pixel size yields an increase in RMS centering error, with the distribution spreading to approximately 8 mrad for the most generous error tolerance and 4:1 camera pixel binning (Figure \ref{fig:pixel_tol}b,c). 

\begin{figure}
\centering
\includegraphics[width=3.2in]{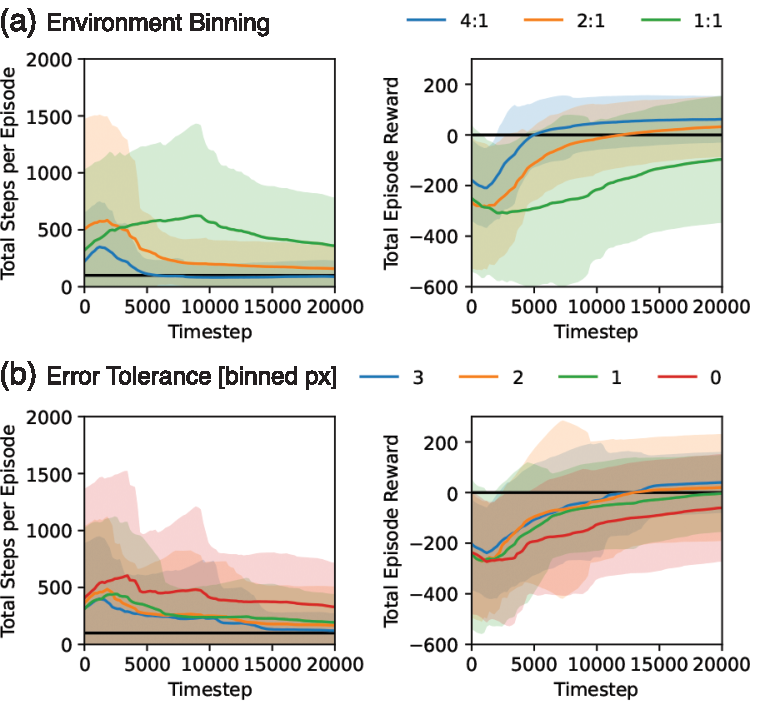}
\caption[name]{Total steps per episode and total episode reward for a) different environment binning and b) different environment goal tolerances. Each line in a) represents the average over 60 experiments while each line in b) represents the average over 80 experiments, and the shaded area shows one standard deviation. A clearer distinction in performance is seen by varying the number of observable states. The black horizontal lines indicates an episode length of 100 and a positive reward threshold for the episode length and reward curves, respectively.}
\label{fig:tol_vs_sub}
\end{figure}

As may be intuitively expected, achieving tighter tolerances in an exponentially larger state space has consequences for convergence time. A longer time to convergence of the RL algorithm is observed for decreasing binning (i.e.~more pixels) of the environment space, with correspondingly lower rewards as well (Figure \ref{fig:tol_vs_sub}a). This can be directly attributed to the many more samples needed to thoroughly explore and form an optimal policy, and further training beyond 20,000 timesteps may address this issue. Perhaps counter-intuitively though, given a degree of binning, there was no clear or consistent relationship between convergence time and goal error tolerance, except for the tightest threshold of 0 pixels (Figure \ref{fig:tol_vs_sub}b). We attribute these results to the size of the observation space, which is related to the degree of exploration necessary to form an optimal policy. In contrast with the exponential decrease in number of states associated with increased binning, a larger goal tolerance decreases the number of non-goal states by only those few within the tolerated range of distances. Additional ambiguity in the {\color{red} effect of these hyperparameters on learning performance across different experiments may be in part due to the stochastic nature of RL, which presents a continual difficulty for reproducing RL behavior despite knowledge of hyperparameters \citep{drl_matters, islam_reproducibility, Lynnerup_reproducibility}. With the current example and set of experiments}, however, the trade-off between run time and error influenced by goal tolerance appears to be minimal compared to the effect of environment binning. 

We highlight that the environment defines the observed position of the bright-field disk as a difference with respect the goal, and the stepwise reward is calculated based on this difference as well. Thus, similar performance should occur for any arbitrary goal position, e.g.~detector center, assuming the lens response is approximately linear over the total shift range. 
If alignment is performed at finer camera pixel size (in mrad), the algorithm precision may also be improved without significant changes in the RL algorithm. At too fine of a sampling interval (or large observed step size from an action), the actual action to alignment shift mapping will exceed the bounds of any pre-trained action mappings, and measurement of the center of mass may be problematic given the size of the bright-field disk on the camera. In this case, adjustments to the scaling of the microscope diffraction shifts or measurement of the center of mass will be necessary to avoid immediately driving the beam off screen. 


\subsection{Reward Function Design}

During learning, the RL algorithm observes only the rewards encountered, rather than the explicit reward function, making calculation of the optimal policy based solely on feedback received from interacting with the environment. As a result, shaping the reward has been shown to have a significant impact on successful learning \citep{Sutton_Barto_2018}. In addition, the performance of the SAC algorithm is shown to be sensitive to proper reward weighting in context with the action and observation space \citep{sachaarnoja}.

\begin{figure*}
\centering
\includegraphics[width=\textwidth]{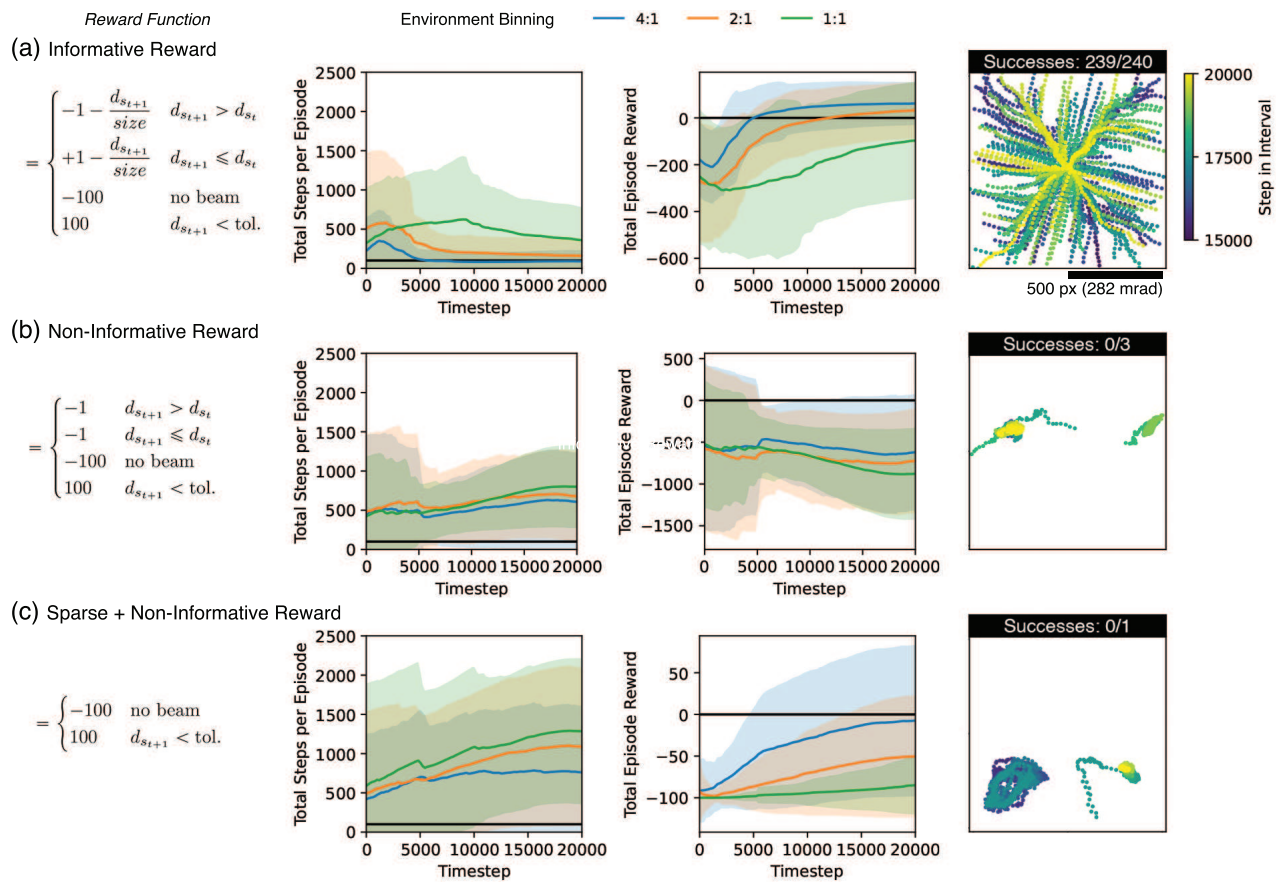}
\caption[name]{Average episode length and episode reward for (a) informative reward (b) non-informative reward and (c) sparse and non-informative reward environments. The reward function is provided at the left. The black horizontal lines indicates an episode length of 100 and a positive reward threshold for the episode length and reward curves, respectively. The final 5,000 observed states and number of successes out of the total completed episodes in this interval during training is shown for each reward function.}
\label{fig:reward_shaping}
\end{figure*}

From tests in the simulated alignment environment, non-informative or sparse rewards decreased the likelihood of successful convergence during 20,000 timestep learning periods (Figures \ref{fig:reward_shaping}b,c) compared to a more informative reward (Figure \ref{fig:reward_shaping}a. Without appropriately weighted rewards, the algorithm did not reliably converge to the goal state within the allotted timesteps. A stepwise reward with a distance-based penalty resulted in a higher average reward and lower average episode length over all tested hyperparameters. Without a distance-based discount of the reward, the network frequently accrued a high negative reward within the learning time by driving the beam off screen or failing to discover the goal state (Figure \ref{fig:reward_shaping}b. Additionally, keeping the lack of distance-based feedback but also removing the stepwise penalty led to longer average episode lengths, with the agent rarely discovering the goal state using this sparse and non-informative reward function (Figure \ref{fig:reward_shaping}c. Drastic reductions in the number of total completed episodes and successful alignments are seen as the reward function is altered, even for the final 15,000-20,000 timestep interval in which a typical agent using the optimized reward function has achieved maximum performance (Figures \ref{fig:genRLcurve} and \ref{fig:reward_shaping}a).

In these three environments, we observe the strong influence that reward shaping has on efficient agent learning, even with only two actions and a bounded two-dimensional observation area. Applying RL to more complex electron microscopy tasks involving higher-dimension action and observation spaces {\color{red} and high-latency controls (such as mechanical stage movement or tilt)} will require careful tuning of the reward function, both to expedite convergence and also avoid issues such as reward hacking {\color{red} and to account for balancing a complex number of parameters in the environment \cite{aisafety, deepmindtokamak}. Even with this consideration, these results indicate that with offline tuning and optimization, an RL agent and environment can be designed to consistently reach a goal state. In situations where control imprecision is compounded with significant state measurement error, the RL algorithm's performance will be detrimentally affected \citep{McKenzie_McDonnell_2019}. More complex algorithms or more robust microscope state measurements could address these potential problems, effectively reducing the challenge of automation to one of general microscope state measurement. }

\subsection{Re-Learning Performance}

\begin{figure}
\centering
\includegraphics[width=3.2in]{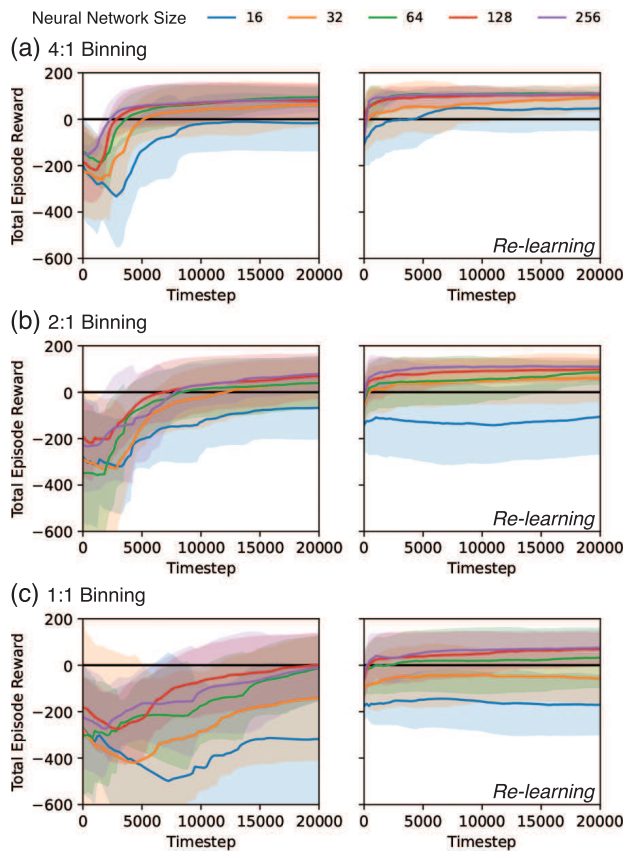}
\caption[name]{Average episode reward during initial learning and re-learning for a) 4:1 b) 2:1 and c) 1:1 binned simulated environments based on specified hidden layer sizes. The black horizontal lines indicates a positive reward threshold.}
\label{fig:relearn}
\end{figure}

Part of the proposed benefit of reinforcement learning-based control algorithms is the ability to re-adjust to changing environments automatically without requiring manual tuning of scaling or other parameters. {\color{red} Two common changes to the bright-field disk for diffraction alignment are convergence angle of the probe and camera length. Since the center of mass is used to measure the position of the disk, the convergence angle has a negligible effect on the measured response (in the absence of a sample). In contrast, changes to camera length essentially magnifies or de-magnifies the reciprocal space information projected onto the phosphor screen. This results in a multiplicative scaling of the actions with respect to other camera lengths.} As such, the re-learning performance of these agents was evaluated for new environments with a different step-size action transformation. The results of these tests, in which the scaling between actions and the observed change in position is reduced by $25\%$, are shown in Figure \ref{fig:relearn}. Across pre-trained models using a variety of hyperparameters, the reinforcement learning agent was able to re-converge to the goal much faster than during initial learning, in many cases in fewer than 500-1,000 timesteps. 

Changing certain hyperparameters, especially the size of hidden layers within the neural network of the RL agent, also affects the performance of re-learning. Agents using artificial neural networks with two hidden layers of sizes 16, 32, 64, 128, and 256 neurons were evaluated for initial learning and re-learning. For environments with larger observation spaces, such as those with less binning, a corresponding increase in the convergence time was observed across all layer sizes, and agents with more hidden nodes generally converged faster than smaller ones, shown in Figure \ref{fig:relearn}. The same trend is seen when the environment step size is adjusted and agents undergo re-learning with this different response. Agents with a larger neural network size perform better in terms of achieving maximal reward as well as accelerated convergence time. Due to the high reward and rapid convergence of re-learning, overfitting to the environment and RL objective are not likely to have occurred. In some cases, such as in agents with hidden layer sizes of 16 or 32 neurons, agents on average did not converge to a positive reward. Thus, influence of the neural network size is significant in solving the task. 

\subsection{Evaluation on the Microscope}

To validate the RL approach and the simulated environment design, an agent with high-performing hyperparameters (reward function, binning, goal error tolerance) was selected for online learning. {\color{red} To prevent poor initializations near the edge of the screen, a more limited starting condition range was given}. In terms of environment response, other differences can be expected by transferring from the simulated environment to the microscope, such as changes in action noise, observation noise, and rotation axes, for which no compensation or changes were made. However, the representation of the beam position in the simulated environment remained unchanged from that of the microscope environment. In other words, the criterion for successful alignment is identical, with the only variation due to center of mass measurement precision. 

\begin{figure}
\centering
\includegraphics[width=3.2in]{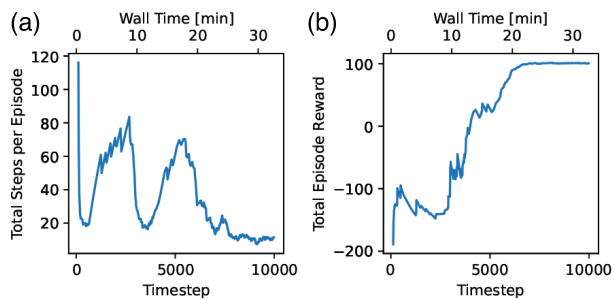}
\caption[name]{{\color{red} Performance curves showing the (a) total steps per episode and (b) total episode reward as a function of elapsed timesteps (and corresponding wall time) during on-microscope learning. } }
\label{fig:RLmicroscope}
\end{figure}

The results of reinforcement learning on the microscope show convergence within 100-200 episodes or about {\color{red} 5,000-6,000 } timesteps, with the agent attaining a minimum number of steps per episode and the expected maximum rewards by the end of learning even in a drastically altered environment (Figure \ref{fig:RLmicroscope}a,b). A more complex curve profile is seen compared to that from learning in the simulated environment as in Figure \ref{fig:genRLcurve}. Several decreases and increases in episode length are seen, associated with either aimless disk movements or movements off-screen. Even so, the reward reaches a minimum quickly and steadily increases within {\color{red} 6,000 timesteps. Looking purely at the number of timesteps, or actions performed, necessary for convergence of the algorithm, gives a value $\approx$ 1,000-2,000 more for the on-microscope agent compared to that of the off-microscope.} Changes in many different aspects of the environment response likely explain the longer adjustment period required for this environment compared to those seen from re-learning with only a scaling change. 

{\color{red} The on-microscope agent operated at an average speed of 5.5 frames or corrections per second, which was primarily limited by a total of about 170 ms of built-in latency for diffraction shift lens response, resets, and exposure time. This yielded a total time to convergence of about 20 minutes, understandably slower than algorithms in the simulated environment that did not account for these factors.} Ultimately, the proof-of-concept approach outlined above shows that simplified environments can be effectively {\color{red} and rapidly} used to tune reinforcement learning agents robust to changes in environment response, offering a calibration-less approach to microscope control.


{\color{red} Based on the current approach, a} future augmented microscope may consist of modular blocks that can be paired together. For example, and depending on the nature of the experiment, an AI algorithm to find a region-of-interest can inform the microscope control to move to a particular location and a control agent can reproduce the conditions for repeated sample measurements for large dataset collection. {\color{red} In the case of high-complexity tasks, such as full column or sample alignment, careful reward function tuning, environment design, and extensive training will  be necessary for achieving desirable results \cite{drl_matters}.} Ultimately, task-based functionality can significantly improve the efficiency of the electron microscope while preserving the flexibility needed to characterize a vast range of materials. {\color{red} That being said, microscope control APIs and documentation are currently not of consistent function or quality across OEMs, which remains an impediment to fully autonomous microscopy. }

\section{Conclusion}

The results of this study demonstrated an approach to design, train, and apply RL methods to augment the electron microscopy workflow. By designing a case study environment and implementing a model for diffraction alignment, we showed successful training without prior knowledge of action space behavior. Further, we showed that by employing offline, simplified environments, both environment design and hyperparameters can be optimized beforehand and successfully validated through deployment on the microscope. {\color{red} By interacting with the microscope using the same feeds and tools available to a human user, RL algorithms demonstrate promise in automating these instruments. } Additionally, we outlined important considerations for the design and optimization of environments and algorithms, offering a starting point for microscopists seeking to augment their own workflows using this approach. In all, this study offers an exploration of how machine learning tools can be designed and utilized to enhance the modern electron microscope workflows.

\section{Acknowledgements}

The authors acknowledge support from the Air Force Office of Scientific Research (FA9550-19-1-0358) and the MIT Research Support Committee. This work made use of the MIT.nano Characterization Facilities. 


\normalsize

\bibliographystyle{MandM}
\bibliography{references}

\begin{thebibliography}{48}
\providecommand{\natexlab}[1]{#1}
\providecommand{\url}[1]{\texttt{#1}}
\providecommand{\urlprefix}{URL }

\bibitem[{Akers et~al.(2021)Akers, Kautz, Trevino-Gavito, Olszta, Matthews,
  Wang, Du \& Spurgeon}]{akersnature}
\textbf{Akers, S., Kautz, E., Trevino-Gavito, A., Olszta, M., Matthews, B.E.,
  Wang, L., Du, Y. \& Spurgeon, S.R.} (2021). Rapid and flexible segmentation
  of electron microscopy data using few-shot machine learning, \textit{npj
  Computational Materials} \textbf{7}, 1–9.

\bibitem[{Amodei et~al.(2016)Amodei, Olah, Steinhardt, Christiano, Schulman \&
  Mané}]{aisafety}
\textbf{Amodei, D., Olah, C., Steinhardt, J., Christiano, P., Schulman, J. \&
  Mané, D.} (2016). Concrete problems in ai safety, \textit{arXiv:160606565} .

\bibitem[{Brockman et~al.(2016)Brockman, Cheung, Pettersson, Schneider,
  Schulman, Tang \& Zaremba}]{openaigym}
\textbf{Brockman, G., Cheung, V., Pettersson, L., Schneider, J., Schulman, J.,
  Tang, J. \& Zaremba, W.} (2016). {OpenAI Gym}, \textit{arXiv:160601540} .

\bibitem[{Callaway(2020)}]{callawayCRYO}
\textbf{Callaway, E.} (2020). {Revolutionary cryo-EM is taking over structural
  biology}, \textit{Nature} \textbf{578}, 201--201.

\bibitem[{Degrave et~al.(2022)Degrave, Felici, Buchli, Neunert, Tracey,
  Carpanese, Ewalds, Hafner, Abdolmaleki, de~las Casas, Donner, Fritz,
  Galperti, Huber, Keeling, Tsimpoukelli, Kay, Merle, Moret, Noury, Pesamosca,
  Pfau, Sauter, Sommariva, Coda, Duval, Fasoli, Kohli, Kavukcuoglu, Hassabis \&
  Riedmiller}]{deepmindtokamak}
\textbf{Degrave, J., Felici, F., Buchli, J., Neunert, M., Tracey, B.,
  Carpanese, F., Ewalds, T., Hafner, R., Abdolmaleki, A., de~las Casas, D.,
  Donner, C., Fritz, L., Galperti, C., Huber, A., Keeling, J., Tsimpoukelli,
  M., Kay, J., Merle, A., Moret, J.M., Noury, S., Pesamosca, F., Pfau, D.,
  Sauter, O., Sommariva, C., Coda, S., Duval, B., Fasoli, A., Kohli, P.,
  Kavukcuoglu, K., Hassabis, D. \& Riedmiller, M.} (2022). Magnetic control of
  tokamak plasmas through deep reinforcement learning, \textit{Nature}
  \textbf{602}, 414–419.

\bibitem[{Ede(2021{\natexlab{a}})}]{partial_scanning}
\textbf{Ede, J.M.} (2021{\natexlab{a}}). {Adaptive partial scanning
  transmission electron microscopy with reinforcement learning},
  \textit{Machine Learning: Science and Technology} \textbf{2}, 045011.

\bibitem[{Ede(2021{\natexlab{b}})}]{Ede_2021deeplearning}
\textbf{Ede, J.M.} (2021{\natexlab{b}}). Deep learning in electron microscopy,
  \textit{Machine Learning: Science and Technology} \textbf{2}, 011004.

\bibitem[{Findlay \& LeBeau(2013)}]{Findlay_LeBeau_2013_detector}
\textbf{Findlay, S. \& LeBeau, J.} (2013). Detector non-uniformity in scanning
  transmission electron microscopy, \textit{Ultramicroscopy} \textbf{124},
  52–60.

\bibitem[{Ge et~al.(2020)Ge, Su, Zhao \& Su}]{deeplearningmatscireview}
\textbf{Ge, M., Su, F., Zhao, Z. \& Su, D.} (2020). {Deep learning analysis on
  microscopic imaging in materials science}, \textit{Materials Today Nano}
  \textbf{11}, 100087.

\bibitem[{Gunawan et~al.(2011)Gunawan, Mkhoyan, Wills, Thomas \&
  Norris}]{gunawan2011dopant}
\textbf{Gunawan, A.A., Mkhoyan, K.A., Wills, A.W., Thomas, M.G. \& Norris,
  D.J.} (2011). Imaging “invisible” dopant atoms in semiconductor
  nanocrystals, \textit{Nano Letters} \textbf{11}, 5553–5557.

\bibitem[{Haarnoja et~al.(2018{\natexlab{a}})Haarnoja, Zhou, Abbeel \&
  Levine}]{sachaarnoja}
\textbf{Haarnoja, T., Zhou, A., Abbeel, P. \& Levine, S.} (2018{\natexlab{a}}).
  {Soft Actor-Critic: Off-Policy Maximum Entropy Deep Reinforcement Learning
  with a Stochastic Actor}, \textit{arXiv:180101290} .

\bibitem[{Haarnoja et~al.(2018{\natexlab{b}})Haarnoja, Zhou, Hartikainen,
  Tucker, Ha, Tan, Kumar, Zhu, Gupta, Abbeel \& et~al.}]{Haarnoja2018}
\textbf{Haarnoja, T., Zhou, A., Hartikainen, K., Tucker, G., Ha, S., Tan, J.,
  Kumar, V., Zhu, H., Gupta, A., Abbeel, P. \& et~al.} (2018{\natexlab{b}}).
  Soft actor-critic algorithms and applications, \textit{arXiv:181205905} .

\bibitem[{Han et~al.(2021)Han, Jang, Cha, Lee, Chung, Jeong, Kim, Chae, Kim,
  Jun, Hwang, Lee \& Ye}]{deeptomoedx}
\textbf{Han, Y., Jang, J., Cha, E., Lee, J., Chung, H., Jeong, M., Kim, T.G.,
  Chae, B.G., Kim, H.G., Jun, S., Hwang, S., Lee, E. \& Ye, J.C.} (2021). {Deep
  learning STEM-EDX tomography of nanocrystals}, \textit{Nature Machine
  Intelligence} \textbf{3}, 267--274.

\bibitem[{Harris(2022)}]{stempy}
\textbf{Harris, C.} (2022). Openchemistry/stempy: stempy 3.0.0,
  \url{https://doi.org/10.5281/zenodo.6546416}.

\bibitem[{He et~al.(2021)He, Wang, Shi, Yang, Li, Wu, Yin \&
  Jin}]{surface_strain_NP}
\textbf{He, T., Wang, W., Shi, F., Yang, X., Li, X., Wu, J., Yin, Y. \& Jin,
  M.} (2021). {Mastering the surface strain of platinum catalysts for efficient
  electrocatalysis}, \textit{Nature} \textbf{598}, 76--81.

\bibitem[{Henderson et~al.(2018)Henderson, Islam, Bachman, Pineau, Precup \&
  Meger}]{drl_matters}
\textbf{Henderson, P., Islam, R., Bachman, P., Pineau, J., Precup, D. \& Meger,
  D.} (2018). Deep reinforcement learning that matters, \textit{Proceedings of
  the AAAI conference on artificial intelligence}, vol.~32.

\bibitem[{Islam et~al.(2017)Islam, Henderson, Gomrokchi \&
  Precup}]{islam_reproducibility}
\textbf{Islam, R., Henderson, P., Gomrokchi, M. \& Precup, D.} (2017).
  Reproducibility of benchmarked deep reinforcement learning tasks for
  continuous control, \textit{arXiv:170804133} .

\bibitem[{Kober et~al.(2013)Kober, Bagnell \&
  Peters}]{Kober_Bagnell_Peters_2013}
\textbf{Kober, J., Bagnell, J.A. \& Peters, J.} (2013). Reinforcement learning
  in robotics: A survey, \textit{The International Journal of Robotics
  Research} \textbf{32}, 1238–1274.

\bibitem[{Kumar et~al.(2021)Kumar, Baker, Bowes, Cabral, Zhang, Dickey, Irving
  \& LeBeau}]{kumar2021pmn}
\textbf{Kumar, A., Baker, J.N., Bowes, P.C., Cabral, M.J., Zhang, S., Dickey,
  E.C., Irving, D.L. \& LeBeau, J.M.} (2021). Atomic-resolution electron
  microscopy of nanoscale local structure in lead-based relaxor ferroelectrics,
  \textit{Nature Materials} \textbf{20}, 62–67.

\bibitem[{LeBeau et~al.(2020)LeBeau, Kumar \&
  Hauwiller}]{LeBeau_Kumar_Hauwiller_2020}
\textbf{LeBeau, J., Kumar, A. \& Hauwiller, M.} (2020). A universal scripting
  engine for transmission electron microscopy, \textit{Microscopy and
  Microanalysis} \textbf{26}, 2958–2959.

\bibitem[{LeBeau \& Stemmer(2008)}]{quant_detectorADF}
\textbf{LeBeau, J.M. \& Stemmer, S.} (2008). {Experimental quantification of
  annular dark-field images in scanning transmission electron microscopy},
  \textit{Ultramicroscopy} \textbf{108}, 1653--1658.

\bibitem[{Lee et~al.(2021)Lee, Nam, Kim, Kim, Lee \& Yoo}]{SEM_RL_autofocus}
\textbf{Lee, W., Nam, H.S., Kim, Y.G., Kim, Y.J., Lee, J.H. \& Yoo, H.} (2021).
  Robust autofocusing for scanning electron microscopy based on a dual deep
  learning network, \textit{Scientific Reports} \textbf{11}, 20933.

\bibitem[{Li et~al.(2019)Li, Sheng \& Ma}]{Li_Sheng_Ma_2019}
\textbf{Li, Q.J., Sheng, H. \& Ma, E.} (2019). Strengthening in multi-principal
  element alloys with local-chemical-order roughened dislocation pathways,
  \textit{Nature Communications} \textbf{10}, 3563.

\bibitem[{Lillicrap et~al.(2015)Lillicrap, Hunt, Pritzel, Heess, Erez, Tassa,
  Silver \& Wierstra}]{lillicrap_continuousRL}
\textbf{Lillicrap, T.P., Hunt, J.J., Pritzel, A., Heess, N., Erez, T., Tassa,
  Y., Silver, D. \& Wierstra, D.} (2015). Continuous control with deep
  reinforcement learning, \textit{arXiv:150902971} .

\bibitem[{Lin et~al.(2021)Lin, Zhang, Wang, Yang \& Xin}]{temimagenet}
\textbf{Lin, R., Zhang, R., Wang, C., Yang, X.Q. \& Xin, H.L.} (2021).
  {TEMImageNet training library and AtomSegNet deep-learning models for
  high-precision atom segmentation, localization, denoising, and deblurring of
  atomic-resolution images}, \textit{Scientific Reports} \textbf{11}, 5386.

\bibitem[{Lynnerup et~al.(2019)Lynnerup, Nolling, Hasle \&
  Hallam}]{Lynnerup_reproducibility}
\textbf{Lynnerup, N.A., Nolling, L., Hasle, R. \& Hallam, J.} (2019). A survey
  on reproducibility by evaluating deep reinforcement learning algorithms on
  real-world robots, \textit{arXiv:190903772} .

\bibitem[{Mastronarde(2005)}]{serialem}
\textbf{Mastronarde, D.N.} (2005). Automated electron microscope tomography
  using robust prediction of specimen movements, \textit{Journal of Structural
  Biology} \textbf{152}, 36–51.

\bibitem[{McKenzie \& McDonnell(2019)}]{McKenzie_McDonnell_2019}
\textbf{McKenzie, M. \& McDonnell, M.D.} (2019). Degradation of performance in
  reinforcement learning with state measurement uncertainty, \textit{2019
  Military Communications and Information Systems Conference (MilCIS)}, 1–5.

\bibitem[{Navalón \& García(2016)}]{np_catalysis}
\textbf{Navalón, S. \& García, H.} (2016). Nanoparticles for catalysis,
  \textit{Nanomaterials} \textbf{6}.

\bibitem[{{Nion Swift}(2022)}]{nionswift}
\textbf{{Nion Swift}} (2022). {Nion Swift},
  \url{https://github.com/nion-software/nionswift}, accessed: 2022-05-22.

\bibitem[{Olszta et~al.(2021)Olszta, Hopkins, Fiedler, Oostrom, Akers \&
  Spurgeon}]{Olszta_sparseauto}
\textbf{Olszta, M., Hopkins, D., Fiedler, K.R., Oostrom, M., Akers, S. \&
  Spurgeon, S.R.} (2021). An automated scanning transmission electron
  microscope guided by sparse data analytics, \textit{arXiv:210914772} .

\bibitem[{{PyJEM}(2022)}]{pyjem}
\textbf{{PyJEM}} (2022). {PyJEM}, \url{https://github.com/PyJEM/PyJEM},
  accessed: 2022-05-22.

\bibitem[{Raffin et~al.(2019)Raffin, Hill, Ernestus, Gleave, Kanervisto \&
  Dormann}]{Stable-Baselines3}
\textbf{Raffin, A., Hill, A., Ernestus, M., Gleave, A., Kanervisto, A. \&
  Dormann, N.} (2019). Stable baselines3,
  \url{https://github.com/DLR-RM/stable-baselines3}, accessed: 2022-05-22.

\bibitem[{Ragone et~al.(2022)Ragone, Saray, Long, Shahbazian-Yassar, Mashayek
  \& Yurkiv}]{Ragone_DLHEAFCN_2022}
\textbf{Ragone, M., Saray, M.T., Long, L., Shahbazian-Yassar, R., Mashayek, F.
  \& Yurkiv, V.} (2022). Deep learning for mapping element distribution of
  high-entropy alloys in scanning transmission electron microscopy images,
  \textit{Computational Materials Science} \textbf{201}, 110905.

\bibitem[{Roccapriore et~al.(2021)Roccapriore, Kalinin \& Ziatdinov}]{autoDKL}
\textbf{Roccapriore, K.M., Kalinin, S.V. \& Ziatdinov, M.} (2021). Physics
  discovery in nanoplasmonic systems via autonomous experiments in scanning
  transmission electron microscopy, \textit{arXiv:210803290} .

\bibitem[{Silver et~al.(2016)Silver, Huang, Maddison, Guez, Sifre, Driessche,
  Schrittwieser, Antonoglou, Panneershelvam, Lanctot, Dieleman, Grewe, Nham,
  Kalchbrenner, Sutskever, Lillicrap, Leach, Kavukcuoglu, Graepel \&
  Hassabis}]{silver_go}
\textbf{Silver, D., Huang, A., Maddison, C.J., Guez, A., Sifre, L., Driessche,
  G.v.d., Schrittwieser, J., Antonoglou, I., Panneershelvam, V., Lanctot, M.,
  Dieleman, S., Grewe, D., Nham, J., Kalchbrenner, N., Sutskever, I.,
  Lillicrap, T., Leach, M., Kavukcuoglu, K., Graepel, T. \& Hassabis, D.}
  (2016). {Mastering the game of Go with deep neural networks and tree search},
  \textit{Nature} \textbf{529}, 484--489.

\bibitem[{Spurgeon et~al.(2021)Spurgeon, Ophus, Jones, Petford-Long, Kalinin,
  Olszta, Dunin-Borkowski, Salmon, Hattar, Yang, Sharma, Du, Chiaramonti,
  Zheng, Buck, Kovarik, Penn, Li, Zhang, Murayama \&
  Taheri}]{Spurgeon_Ophus_datadriven}
\textbf{Spurgeon, S.R., Ophus, C., Jones, L., Petford-Long, A., Kalinin, S.V.,
  Olszta, M.J., Dunin-Borkowski, R.E., Salmon, N., Hattar, K., Yang, W.C.D.,
  Sharma, R., Du, Y., Chiaramonti, A., Zheng, H., Buck, E.C., Kovarik, L.,
  Penn, R.L., Li, D., Zhang, X., Murayama, M. \& Taheri, M.L.} (2021). Towards
  data-driven next-generation transmission electron microscopy, \textit{Nature
  Materials} \textbf{20}, 274–279.

\bibitem[{Sutton \& Barto(2018)}]{Sutton_Barto_2018}
\textbf{Sutton, R.S. \& Barto, A.G.} (2018). \textit{Reinforcement learning: An
  introduction}, Adaptive computation and machine learning series, The MIT
  Press, second ed.

\bibitem[{Tegunov \& Cramer(2019)}]{tegunov_warp}
\textbf{Tegunov, D. \& Cramer, P.} (2019). Real-time cryo-electron microscopy
  data preprocessing with warp, \textit{Nature Methods} \textbf{16},
  1146–1152.

\bibitem[{{Thermo Fisher Scientific}(2022)}]{epu2}
\textbf{{Thermo Fisher Scientific}} (2022). {EPU 2 EM Software},
  \urlprefix\url{https://www.thermofisher.com/us/en/home/electron-microscopy/products/software-em-3d-vis/epu-software.html}.

\bibitem[{Uusimaeki et~al.(2019)Uusimaeki, Wagner, Lipinski \& Kaegi}]{autoEM}
\textbf{Uusimaeki, T., Wagner, T., Lipinski, H.G. \& Kaegi, R.} (2019).
  {AutoEM: a software for automated acquisition and analysis of nanoparticles},
  \textit{Journal of Nanoparticle Research} \textbf{21}, 122.

\bibitem[{Vasudevan et~al.(2021)Vasudevan, Ghosh, Ziatdinov \&
  Kalinin}]{vasudevan_ebeam}
\textbf{Vasudevan, R.K., Ghosh, A., Ziatdinov, M. \& Kalinin, S.V.} (2021).
  {Exploring Electron Beam Induced Atomic Assembly via Reinforcement Learning
  in a Molecular Dynamics Environment}, \textit{arXiv:210411635} .

\bibitem[{Vinyals et~al.(2019)Vinyals, Babuschkin, Czarnecki, Mathieu, Dudzik,
  Chung, Choi, Powell, Ewalds, Georgiev, Oh, Horgan, Kroiss, Danihelka, Huang,
  Sifre, Cai, Agapiou, Jaderberg, Vezhnevets, Leblond, Pohlen, Dalibard,
  Budden, Sulsky, Molloy, Paine, Gulcehre, Wang, Pfaff, Wu, Ring, Yogatama,
  Wünsch, McKinney, Smith, Schaul, Lillicrap, Kavukcuoglu, Hassabis, Apps \&
  Silver}]{starcraft}
\textbf{Vinyals, O., Babuschkin, I., Czarnecki, W.M., Mathieu, M., Dudzik, A.,
  Chung, J., Choi, D.H., Powell, R., Ewalds, T., Georgiev, P., Oh, J., Horgan,
  D., Kroiss, M., Danihelka, I., Huang, A., Sifre, L., Cai, T., Agapiou, J.P.,
  Jaderberg, M., Vezhnevets, A.S., Leblond, R., Pohlen, T., Dalibard, V.,
  Budden, D., Sulsky, Y., Molloy, J., Paine, T.L., Gulcehre, C., Wang, Z.,
  Pfaff, T., Wu, Y., Ring, R., Yogatama, D., Wünsch, D., McKinney, K., Smith,
  O., Schaul, T., Lillicrap, T., Kavukcuoglu, K., Hassabis, D., Apps, C. \&
  Silver, D.} (2019). {Grandmaster level in StarCraft II using multi-agent
  reinforcement learning}, \textit{Nature} \textbf{575}, 350--354.

\bibitem[{Yang et~al.(2021)Yang, Choi, Cho, Agyapong‐Fordjour, Park, Yun,
  Kim, Han, Lee, Kim \& Kim}]{yang_DL_dopant2021}
\textbf{Yang, S., Choi, W., Cho, B.W., Agyapong‐Fordjour, F.O., Park, S.,
  Yun, S.J., Kim, H., Han, Y., Lee, Y.H., Kim, K.K. \& Kim, Y.} (2021). Deep
  learning‐assisted quantification of atomic dopants and defects in 2d
  materials, \textit{Advanced Science} \textbf{8}, 2101099.

\bibitem[{Yin et~al.(2020)Yin, Brittain, Borseth, Scott, Williams, Perkins,
  Own, Murfitt, Torres, Kapner, Mahalingam, Bleckert, Castelli, Reid, Lee,
  Graham, Takeno, Bumbarger, Farrell, Reid \& da~Costa}]{petascale}
\textbf{Yin, W., Brittain, D., Borseth, J., Scott, M.E., Williams, D., Perkins,
  J., Own, C.S., Murfitt, M., Torres, R.M., Kapner, D., Mahalingam, G.,
  Bleckert, A., Castelli, D., Reid, D., Lee, W.C.A., Graham, B.J., Takeno, M.,
  Bumbarger, D.J., Farrell, C., Reid, R.C. \& da~Costa, N.M.} (2020). A
  petascale automated imaging pipeline for mapping neuronal circuits with
  high-throughput transmission electron microscopy, \textit{Nature
  Communications} \textbf{11}, 4949.

\bibitem[{Ziatdinov et~al.(2017)Ziatdinov, Dyck, Maksov, Li, Sang, Xiao,
  Unocic, Vasudevan, Jesse \&
  Kalinin}]{Ziatdinov_defects_transformsseg_Kalinin_2017}
\textbf{Ziatdinov, M., Dyck, O., Maksov, A., Li, X., Sang, X., Xiao, K.,
  Unocic, R.R., Vasudevan, R., Jesse, S. \& Kalinin, S.V.} (2017). Deep
  learning of atomically resolved scanning transmission electron microscopy
  images: Chemical identification and tracking local transformations,
  \textit{ACS Nano} \textbf{11}, 12742–12752.

\bibitem[{Ziatdinov et~al.(2021{\natexlab{a}})Ziatdinov, Ghosh, Wong \&
  Kalinin}]{atomai}
\textbf{Ziatdinov, M., Ghosh, A., Wong, T. \& Kalinin, S.V.}
  (2021{\natexlab{a}}). Atomai: A deep learning framework for analysis of image
  and spectroscopy data in (scanning) transmission electron microscopy and
  beyond, \textit{arXiv:210507485} .

\bibitem[{Ziatdinov et~al.(2021{\natexlab{b}})Ziatdinov, Jesse, Sumpter,
  Kalinin \& Dyck}]{trackingsideep}
\textbf{Ziatdinov, M., Jesse, S., Sumpter, B.G., Kalinin, S.V. \& Dyck, O.}
  (2021{\natexlab{b}}). {Tracking atomic structure evolution during directed
  electron beam induced Si-atom motion in graphene via deep machine learning},
  \textit{Nanotechnology} \textbf{32}, 035703.

\end{thebibliography}

\end{document}